# The Dark Energy Camera (DECam)

K. Honscheid, D. L. DePoy, T. Abbott, J. Annis, M. Antonik, M. Barceló, R. Bernstein, B. Bigelow, D. Brooks, E. Buckley-Geer, J. Campa, L. Cardiel, F. Castander, J. Castilla, H. Cease, S. Chappa, E. Dede, G. Derylo, T. Diehl, P. Doel, J. DeVicente, J. Eiting, J. Estrada, D. Finley, B. Flaugher, E. Gaztanaga, D. Gerdes, M. Gladders, V. Guarino, G. Gutierrez, J. Hamilton, M. Haney, S. Holland,  D. Huffman, I. Karliner, D. Kau, S. Kent, M. Kozlovsky, D. Kubik, K. Kuehn, S. Kuhlmann, K. Kuk, F. Leger, H. Lin, G. Martinez, M. Martinez, W. Merritt, J. Mohr, P. Moore, T. Moore, B. Nord, R. Ogando, J. Olsen, B. Onal, J. Peoples, T. Qian, N. Roe, E. Sanchez, V. Scarpine, R. Schmidt, R. Schmitt, M. Schubnell, K. Schultz, M. Selen, T. Shaw, V. Simaitis, J. Slaughter, C. Smith, H. Spinka, A. Stefanik, W. Stuermer, R. Talaga, G. Tarle, J. Thaler, D. Tucker, A. Walker, S. Worswick, A. Zhao
For the Dark Energy Survey Collaboration

In this paper we describe the Dark Energy Camera (DECam), which will be the primary instrument used in the Dark Energy Survey. DECam will be a 3 sq. deg. mosaic camera mounted at the prime focus of the Blanco 4m telescope at the Cerro-Tololo International Observatory (CTIO). It consists of a large mosaic CCD focal plane, a five element optical corrector, five filters (g,r,i,z,Y), a modern data acquisition and control system and the associated infrastructure for operation in the prime focus cage. The focal plane includes of 62 2K x 4K CCD modules (0.27″/pixel) arranged in a hexagon inscribed within the roughly 2.2 degree diameter field of view and 12 smaller 2K x 2K CCDs for guiding, focus and alignment. The CCDs will be 250 micron thick fully-depleted CCDs that have been developed at the Lawrence Berkeley National Laboratory (LBNL). Production of the CCDs and fabrication of the optics, mechanical structure, mechanisms, and control system for DECam are underway; delivery of the instrument to CTIO is scheduled for 2010.

## 1. INTRODUCTION

The discovery that the universe is *accelerating*, not slowing down from the mass it contains, is the surprise that sets the initial research program of 21st Century cosmology. The Dark Energy Survey (DES) [1] is a next generation sky survey aimed directly at understanding this mystery. Over 525 nights DES will survey a 5000 sq-degrees area of the sky in five filter bands using a new 3 deg$^2$ camera (DECam) on the Blanco 4m telescope [2]. The survey area will overlap with the South Pole Telescope [3] and the VHS VISTA survey [4]. In particular, DES will provide Y band observations for the Vista Hemispherical Survey (VHS), which in exchange will supplements DES with near-infrared measurements that improve photometric redshift precision and accuracy.

DES is designed to measure the dark energy equation of state parameter with four complementary techniques: galaxy cluster counts, weak lensing, angular power spectrum and type Ia supernovae. The combination of these techniques will produce a factor of ~4.6 improvement over current experiments in the figure of merit defined by the Dark Energy Task Force. Details on the DES constraints and systematic uncertainties can be found in several DES whitepapers [5].

## 2. DECAM

We have developed a design for DECam that requires only modest development of proven technologies with the goal of minimizing the construction time and costs while maximizing the scientific potential. Table 1 contains a summary of the DES survey parameters and the expected performance of the CTIO site and DECam. Figure 2 shows a cross section of DECam and a picture of the Blanco telescope.  DECam will replace the entire prime focus cage of the Blanco and attach to the existing spider support fins. The major components of DECam are a ~520 megapixel optical CCD camera with vacuum and cryogenic controls, a compact low noise CCD readout system housed in actively cooled crates, a combination shutter-filter system to house the shutter, the DES filters (g,r,i,z,Y) plus slots for three additional filters that could be provided by the observer community, a modern read-out and control system, and a wide-field optical corrector (2.2 deg. field of view).

Table 1: Expected performance of DECAM, Blanco, and CTIO site

| | |
|---|---|
| Blanco Effective Aperture/ f number @ prime focus | 4 m/ 2.7 |
| Blanco Primary Mirror - 80% encircled energy | 0.25 arcsec |
| Optical Corrector Field of View | 2.2 deg. |
| Corrector Wavelength Sensitivity | <350-1050 nm |
| Filters | SDSS g, r, i, z, Y  (400-1050 nm) |
| Effective Area of CCD Focal Plane | 3.0 sq. deg. |
| Image CCD pixel format/ total # pixels | 2K X 4K/ 520 Mpix |
| Guide, Focus & Alignment Sensor CCD pixel format | 2K X 2K |
| Pixel Size | 0.27 arcsec/ 15 μm |
| Readout Speed/Noise requirement | 250 kpix/sec/ 10 e |
| Survey Area | 5,000 sq. deg. total |
|     SPT overlap | RA -60 to 105, DEC -30 to -65 |
| | RA -75 to -60 , DEC -45 to -65 |
|     SDSS stripe 82 | RA -50 to 50, Dec -1 to 1 |
|     Connection region | RA 20 to 50, Dec -30 to -1 |
| Survey Time/Duration | 525/5 (nights/years) |
| Median Site Seeing  Sept. – Feb. | 0.65 arcsec |
| Median Delivered Seeing with Mosaic II on the Blanco | 0.9-1.0 arcsec (V band) |
| Limiting Magnitude: 10σ in 1.6" aperture assuming 0.9" seeing, AB system | g=24.7, r=24.2, i=24.4, z=23.9 |
| Limiting Magnitude: 5σ for point sources assuming 0.9" seeing , AB system | g=26.1,r=25.6, i=25.8, z=25.3 |

The CCD vessel and corrector are supported as a single unit by a hexapod that will provide adjustability in all degrees of freedom including focus, lateral translations, tip, and tilt. Note that the current prime focus cage of the Blanco can flip such that the F/8 mirror that is mounted to the back of the cage, behind the CCD vessel, points towards the primary mirror for Cassegrain observations. DES will maintain this capability without the flip by attaching the existing F/8 mirror to the front of the new DES prime focus cage when DECam is not in use.

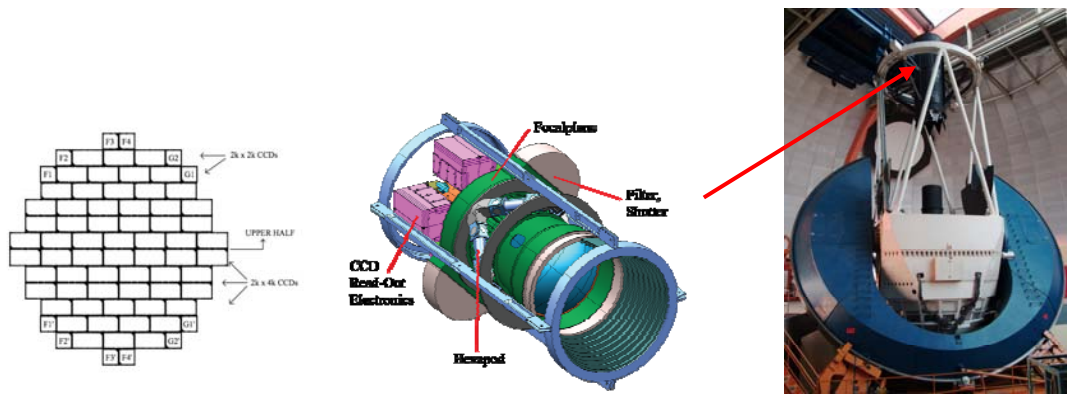

Figure 1: Schematic view of the DECam focal plane, the complete instrument and a picture of the Blanco telescope. The arrow connecting the two shows the position of the Blanco telescope prime focus, which will be the mounting location of the instrument.

The DES survey strategy is based on taking multiple short exposures (100 sec) and adding them together to reach the required depth in each filter pass-band. This strategy minimizes the systematic uncertainties resulting from effects such as atmospheric variations. In the first two years we plan to tile the entire survey area multiple times to enable early scientific results. To maximize the total time available for exposures, we plan to read out the image CCDs while the telescope is slewing to a new position. Current measurements indicate that the time it takes for a typical DES slew of 2 deg. is ~35 sec. Upgrades planned by CTIO for the telescope control system will reduce this to less than 20 sec. At a CCD readout rate of 250 kpix/sec image acquisition will take 17 sec and fit within the telescope slew time.

We plan to read out the alignment and focus CCDs with the image CCDs to provide the possibility for image by image corrections. Currently at the Blanco, observations are interrupted 2-3 times/night to correct the telescope focus and the corrector-primary mirror alignment is checked and adjusted every few months. Experience with DECam on the Blanco will determine the frequency at which corrections are needed to maintain the best image quality. The guide CCDs need to provide signals to the telescope controls system at a rate of ~ 1 Hz. To achieve this rate, exposures of ~ 0.5 sec are envisioned and only a small area centered on the guide star will be read out.

## 2.1. DES Focal plane and CCD Readout

To efficiently obtain $z$-band images for high-redshift ($z\sim1.3$) galaxies, we use fully depleted, high-resistivity, 250 micron thick CCDs [6] that have been designed and developed at LBNL. The thickness of the design has two important implications for DES: fringing is negligible and the QE of these devices is > 50% in the $z$ band, a factor of 5-10 higher than traditional thinned astronomical devices. The read noise of the CCDs is a function of readout speed. At a rate of 250 kpix/sec we see a read noise of ~7 electrons, which is meets the DES requirement of <10 electrons. The DES focal plane, shown in Figure 1, will consist of sixty-two 2K x 4K CCDs (0.27"/pixel) arranged in a hexagon covering an imaging area of 3 sq. degrees. Smaller format CCDs for guiding (G), focusing (F) and alignment (WF) are located at the edges of the focal plane.

The DES CCDs are being packaged and tested at Fermilab, capitalizing on the experience and infrastructure associated with construction of silicon strip detectors for the Fermilab Tevatron program. The four-side buttable package for the 2k x 4k devices builds on techniques developed by LBNL and Lick Observatory. The DES science requires a uniform PSF. This translates into a requirement that each CCD module have a flatness (peak-to-peak) of < 10 microns. Careful matching of the thermal expansion coefficients is required to achieve this level of flatness when the CCD modules are cooled to an operating temperature between -80 C and -120 C. In the focal plane package, called the pedestal package, the front side of the CCD is glued and wire-bonded to an Aluminum Nitride (AlN) circuit board, then to an AlN spacer and an Invar foot. A 37 pin connector is located in the center of the device and is accessed through a hole in the foot. Due to the large number of devices that will need to be packaged and tested, an emphasis on ease of assembly and handling is incorporated into the design.

CCD testing and characterization is currently being performed at Fermilab using the Monsoon readout system developed originally by NOAO [7]. Each DES CCD has two video readout channels and requires connection to 15 clock inputs, and 7 bias voltages. In addition, each CCD has a 4-wire temperature sensor on the AlN circuit board that will be used to track the CCD temperature during operation. We have modified/customized the Monsoon architecture and electronics boards to meet specific DES requirements. A new 12 channel acquisition board was developed and the communication link was upgraded to the S-Link system developed by CERN. We have demonstrated that we can achieve low noise operation at the pixel clock rates required by DES [8]. With these modifications the readout for the DES focal plane will fit into three 10 slot dual backplane crates. The CCD readout crates will be located in thermally controlled housings. These housings will maintain the CCD electronics at a constant temperature to ensure constant gain performance. The outside of the housing will follow the ambient night temperature to avoid creation of thermal plumes in the optical path. A liquid coolant, likely a water alcohol mixture, will be transported to these housings through insulated lines to remove the heat from the electronics.

## 2.2. CCD Camera and Optical Corrector

The camera vessel contains the last element of the optical corrector (C5, see below) and the CCD focal plane array; the design of the imager vessel is nearing completion. The current design uses a re-circulating Liquid Nitrogen (LN$_2$) system that is connected to the focal plane support plate by copper straps. Each strap will have a temperature monitor and a heater that can tune the temperature of the focal plane support plate. The CCD operating temperature will be between -120 deg. C and -80 deg. C and will be determined by study of the quantum efficiency as a function of temperature. Finite element models indicate that with 12 straps the focal plane support plate temperature can be uniform to <2 deg. C across the 525 mm diameter focal plane. The C5 lens is athermally constrained within a cell using radial High Density Polyethylene (HDPE) spacers that are sized to compensate for the CTE difference between the lens and the cell. The C5 cell mounts to the front flange of the camera vessel with an o-ring seal between the lens and the flange. The CCD array is attached to the vessel using bipod style mounts. The bipods mounts thermally isolate the array from the rest of the vessel and locate the array with respect to C5. The CCD vessel has a 61 cm diameter and is approximately 84 cm inches in the axial direction. A full sized prototype camera vessel has been constructed and is being used for cooling tests. It provides a test bed for multi CCD readout in an environment very close to what is imagined for the final CCD vessel design.

The optical corrector design consists of five fused silica lenses that produce an unvignetted 2.2$^{o}$ diameter image area. The largest element (C1) is 930 mm diameter. The last element (C5) is the window on the CCD vacuum vessel. The spacing between elements 3 and 4 allows room for the ~570 mm diameter filters to slide in and out of the optical path. To maximize the stiffness of the corrector barrel the shutter will be combined with the filter changer. Studies indicate that an Atmospheric Dispersion Corrector (ADC) would not significantly improve the optical performance for DES and thus the DECam optical corrector does not include one [8]. The lenses will be mounted into cells and then into the corrector barrel. Detailed design of the cells and the barrel is in progress. A preliminary Finite Element Model (FEM) of the barrel and CCD vessel has been used to estimate the deflections of each of the lenses under gravity loading. Reinforcing tubes have been added around the filter-shutter system with the goal of keeping the deflections of C1 and the focal plane to < 25 microns. The results of the FEM have been fed into a sensitivity analysis of the optics to generate an estimate of the resulting changes to the image quality and the initial indications are that the design is sufficiently stiff. Future work will include a more detailed FEM and will investigate making the barrel and CCD vessel stiffer and lighter.

The DECam shutter is a design developed by Bonn University. It has sliding blades (a "flying slit") for fast exposures and to obtain equal exposure times in both directions of travel. The DES filters will be ~570 mm diameter and 20-30 mm thick. The main challenge in filter fabrication is expected to be uniformity over this large area. Other projects such as PanSTARRS, LSST, the WIYN ODI and the Discovery Channel Telescope have similar requirements. DES has developed a set of detailed filter specifications that has been sent out for budgetary quotes; responses are under review. The filter exchange mechanism will house the five DES filters and up to three additional filters. The filter cartridge and the corrector barrel are purged with dry nitrogen to maintain a constant humidity environment. The cartridge system places the filters at different positions in the optical path. Initial feedback from the filter vendors indicate that the filters may be as thick as 30 mm and with the addition of the mechanical housings this will produce substantial offsets between the filters in different cassettes. The optical beam is roughly a constant diameter in this region and we expect that all the filters will be the same size. Scattered light analysis of the optics and associated mechanical structures (including the telescope) is complete. The analysis revealed several improvements to the telescope and barrel design that should help control scattered/stray light in the camera, which will lead to improved performance.

## 2.3. Alignment and Focus

The design for the support and position adjustment of the camera/barrel assembly in the prime focus cage incorporates a hexapod capable of accurate full 6-axis motion. The base of the hexapod will be mounted to a ring built into the prime focus cage. The corrector will be attached to the top of the hexapod using the front plate of the filter/shutter assembly. Note that the CCD vessel is aligned and bolted to the corrector and the combination camera/barrel system moves as a single unit. The hexapod system will provide both the focus adjustments and the capability to keep the corrector aligned to the primary mirror.

The 2k x 2k focus and alignment CCDs will be mounted either above or below the nominal surface of the focal plane. Comparison of the out-of-focus images above and below the focal plane will indicate the direction of motion needed to bring the image back into focus. The alignment CCDs and a variant of the "donut" method now under development at CTIO for the SOAR telescope will be used to monitor the alignment of the corrector to the primary mirror. The focus and alignment CCDs will be read out with the image CCDs providing image-by-image information on the telescope focus and alignment. We also plan to install a laser-alignment system, which will directly measure the relative positions of the mirror and the corrector without needing to operate the CCDs. Information from one or both of these systems will be used to determine if the corrector needs to be moved laterally in the cage to be better aligned with the primary mirror.

## 2.4. Survey Image System Process Integration (SISPI)

The DES mountain top data acquisition system called the Survey Image System Process Integration (SISPI) is shown schematically in Figure 2. Image data flows from the focal plane CCDs and the Monsoon front end electronics to the Image Acquisition and Image Builder systems before it is recorded on a storage device and handed over to the DES Data Management system. The data flow is coordinated by the Observation Control system (OCS) which determines the exposure sequence and is assisted by the Instrument Control (ICS) and Image Stabilization systems. Pointing information, correction signals derived from the guide CCDs and other telemetry information is exchanged with the Blanco Telescope Control system (TCS). The ICS implements hardware control loops for the shutter, filter changer, cooling and the focal plane heater system. The performance requirements for the DECam read-out and control system are set by the size of the focal plane, the read-out time and the typical exposure time of 100s. With 62 CCDs or 520

Mpixels and 16 bits per pixel the size of a DECam exposure is approximately 1 GByte. At a rate of 250 kpix/s it takes about 17 seconds to transfer the data from the focal plane to the computers of the Image Acquisition system. During this time the telescope slews to the next position. Via Gigabit Ethernet links the image data is shipped to the Image Building system where it is packaged in multi-extension FITS files and stored on a disk. The SISPI applications shown in Figure 2b are built upon a common software infrastructure layer which provides the message passing system, the facility database and support for configuration and initialization.

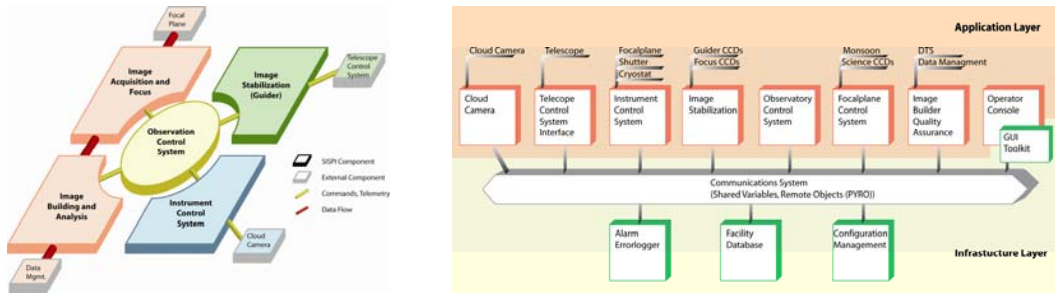

Figure 2: A block diagram of the DES data acquisition system (a) and details of the software architecture (b).

In the design of the SISPI infrastructure and communication software we distinguish between *Command* messages and *Telemetry* data. Commands are used to request information from a remote application or to activate a remote action. The Command or Message Passing system is implemented using a Client-Server design pattern based the CTIO developed Soar Messaging Library (SML). *Telemetry* in this context refers to any information an application provides and that can be of use to another application. In order to provide this functionality we implemented a Shared Variable system (SVE) using the Publish-Subscribe design pattern. Prototype components of the read-out and control system are operational and are being used in ongoing integration test.

## 3. DES PROJECT STATUS AND SCHEDULE

The Dark Energy Survey Collaboration (see www.darkenergysurvey.org) formed in 2004 and began the process of defining the project cost and schedule and securing funding. In May 2008 DECam received CD-2/CD-3a approval from the DOE and full CD-3 approval is expected soon. We are on schedule for delivery of DECam to CTIO at the end of 2010 and anticipate that the first DES observing season will begin in October 2011.